\documentstyle[prl,aps,twocolumn,epsfig]{revtex}

\begin{document}

\title{Magnetotransport measurements on freely suspended two-dimensional 
electron gases}
\draft
\author{
R.H.~Blick$^{* \dagger}$, F.G.~Monzon$^*$, W.~Wegscheider$^{**\ddagger}$, M.~Bichler$^{**}$, \\
F. Stern$^{***\amalg}$, and M.L.~Roukes$^*$
}
 
\address{
$^*$~California Institute of Technology, Condensed Matter Physics 114-36, 
Pasadena, California 91125, USA. \\
$^{**}$~Walter-Schottky-Institut der Technischen Universit\"at M\"unchen, 
Am Coulombwall, 85748 Garching, Germany.        \\
$^{***}$ IBM T.J. Watson Research Center, P.O. Box 218, 
Yorktown Heights, New York 10598, USA.
}

\date{\today}
\maketitle

%\widetext
%\tightenlines

% %%%%%%%%%%%%%%%%%%%%%% Kurzfassung %%%%%%%%%%%%%%%%%%%%%%%%
\begin{abstract}
We present magnetotransport measurements on freely suspended 
two-dimen\-sional electron gases from AlGaAs/GaAs heterostructures. 
The technique to realize such devices relies on a specially MBE grown 
GaAs/AlGaAs-heterostructure including a sacrificial layer.
We fabricated simple Hall-bars as well as quantum cavities and quantum 
dot systems. We find well-pronounced Shubnikov-de Haas oscillations and
observe commensurability resonances, allowing characterization 
of the electron gas in these 100~nm thin membranes. 
\pacs{07.10.Cm,73.50.-h}
\end{abstract}

%\narrowtext

%\newpage

Recently the realization of suspended, monocrystalline, GaAs nanostructures 
containing a three-dimensional electron gas has been
demonstrated~\cite{tighe}. 
Previous work has shown the possibility of using a GaAs high-electron
mobility transistor to achieve sensitive piezoelectric 
detection of strain~\cite{beck}. We are combining these techniques to
enable new means of detecting motion in nanomechanical systems and 
to study interactions in coupled electron-phonon systems of reduced 
dimensionality for both the electrons {\it and} the phonons.
Free-standing structures incorporating a high-mobility electron gas have 
interesting device applications: A high mobility 2DEG system provides a
unique approach to 
implement wideband, extremely sensitive displacement detection.  These
systems also
offer prospects for very sensitive bolometers and represent model systems for 
high sensitivity calorimetry~\cite{schwab}.  One very interesting potential
application is in heat capacity measurements on two-dimensional electron
gases (2DEGs).
The sensitivity of these types of thermal devices is enhanced by their small 
volume, which can be of order of $\sim 3~\mu$m$^3$, significantly smaller
than the usual dimensions of $\sim 10^8~\mu$m$^3$~\cite{tighe,schwab}.
 
In this work we discuss the processing technique of such devices and
present first results on magnetotransport measurements on suspended two- and
zero-dimensional systems. The processing employs a specially-designed
MBE-grown 2DEG
heterostructure in which a sacrificial layer is included. The layer
structure is shown in Fig.~1(a): The structural layer stack, from which
the devices are
formed, comprises seven individual layers having a total thickness of 100~nm. 
Top and bottom are formed by thin GaAs cap layers preventing oxidation
of the AlGaAs:Si donor layers which follow beneath.
The central 15~nm thick GaAs layer forms a quantum well sustaining a high
mobility 2DEG located 37~nm below the top surface and is surrounded by two
AlGaAs spacer layers. Below the structural layer stack
is a 400~nm Al$_{0.8}$Ga$_{0.2}$As sacrificial layer. 

The heterostructure was designed by modelling the conduction band lineup
numerically. These calculations employed a Poisson-Schr\"odinger solver
program, written
by Laux and Kumar, which neglects many-body
effects~\cite{kumar}. In 
contrast to common 2DEG heterostructure configurations, here an additional
GaAs layer is included to avoid deleterious carrier depletion effects from
the lower
surface once it becomes exposed, i.e. after the sacrificial etch. The
calculations
indicate the donor regions are particularly susceptible to illumination,
and that parallel
conduction might be possible. However, in our measurements no obvious
contribution 
from such channels even after continuous illumination was evident as
will be discussed below. Hence,
we only have to consider a single 2DEG layer in the evaluation below. 

To realize these suspended nanostructures with three-dimensional relief we
employ multiple steps of optical and electron beam lithography, followed by a
combination of pattern transfer steps. The latter involve both anisotropic
ion-beam (dry)
and chemically-selective (wet) etching techniques. The first step of the
processing procedure is fabrication of standard Au/Ni/Ge ohmic contacts to
the GaAs.
These require an annealing step after their deposition. The second step is
definition of contacts by optical and electron beam litho\-graphy and
patterning with Ni.
These contacts serve as a metallic etch mask during the subsequent ion-beam
etch step. The Ni etch
mask is selectively
removed by another brief wet etch step employing FeCl$_3$ after the dry
etch. The device geometry is machined by anisotropic, chemically assisted
ion beam etching (CAIBE) with chlorine gas.  The final step then is the
selective
chemical etch to remove the sacrificial layer. Depending on the
stochiometry of the
structural layer, either a 2\% solution of HF or concentrated HCl was
employed~\cite{tighe,blauvelt}. 

A typical freely suspended 2DEG samples is shown in Fig.~1(b):
In this case the device has a shape similar to a conventional Hall bar with
dimensions of $1 \times 2~\mu$m$^2$ (width $\times$ length). In Fig.~2(a) a
scanning 
electron microscope (SEM) micrograph of the same device as in Fig.~1(b) is
shown, but taken 
at a steeper angle; the sacrificial layer is clearly seen to be completely
removed from beneath the structure. In comparison to usual Hall-bars with a
wide 2DEG region and small contact areas we had to choose a simplified design,
since the
underetching effectively limits the lateral extension of the system. For
larger membranes the
mechanical supports are commonly undercut as well, leading to a collapse of
the whole structure. 
Although the contact areas are fairly large compared to the device size,
the mini-Hall-bar allows us to monitor the longitudinal voltage
drop properly.

Our quantum devices, also exemplified in Fig.~2(b), represent the first truly 
suspended quantum dots.  The single dot has a final diameter of 800~nm,
while the diameter of the two 
coupled dots is of order of 400~nm.  As discussed below, the actual {\it
electronic} diameter of these devices is reduced by edge
depletion~\cite{choi}.  In the present
case, coupling between the dots is mediated by the constriction regions
connecting them.  Careful design of the geometry of these regions allows to
control the
carrier depletion and hence the degree of coupling to the leads.
Fabrication of additional gate 
contacts in these regions, not attempted in these preliminary
investigations, would permit definition of variable tunneling barriers. 

At first, low temperature magnetotransport measurements were carried out to
characterize the electron gas in the mini-Hall-bar and the quantum devices.  
Fig.~2(c) displays Shubnikov-de Haas (SdH) oscillations 
in the longitudinal resistance for one of the suspended Hall bar samples
after illumination with a red light emitting diode for 60~sec. Prior to
illumination the
resistance commonly was a factor of five larger, as seen in the values
obtained for the cavities~A and~B in Fig.~3. The reason for
this deviations might be also found in a varying surface tension of the
different suspended 2DEGs. 
The electron density is finally obtained from 
the SdH-oscillations shown in Fig.~2(c) through the standard relation 
$\Delta (1/B) = g_s e / (h n_s)$, where $g_s$ is the spin degeneracy 
factor (here $g_s = 2$), which yields 
$n_s \sim 5.5 \times 10^{15}$ m$^{-2}$. Note that the Landau levels are well
developed even at fairly low fields. In the inset of Fig.~2(c) the $(1/B)$
spectrum of the oscillations for the illuminated shows only a single period.
Hence, a parallel
conduction channel appears to be absent or can at least be neglected for 
the transport data, despite illumination and the model calculations.  
The most likely explanation for this is the fact that lattice strain
effects are
not included in the model calculations, but are expected to play a significant
role for these suspended devices. 

For the mini-Hall-bar the sheet resistance is of 
the order of $\rho = \rho_{xx} \sim 4$~k$\Omega / \Box$ without
illumination, and $\sim 700  \Omega / \Box - 800  \Omega / \Box$ after
illumination.  The values are obtained by extrapolating the resistance
values from $B = 1.5$~T
to $B \rightarrow 0$~T, in order to circumvent contributions by scattering
around $B = 0$ (seen in the strong {\it negative} magnetoresistance). 
As noted before, the length/width ratio of the Hall bars is 
$l / w = 2~\mu$m$ /1~\mu$m$ = 2$.  Combining this zero-field resistance of 
approximately $700  \Omega / \Box$ with the measured carrier density, we 
obtain an electron mobility after illumination of $1.4 - 1.6$~m$^2$/Vs. 
This is a factor of 27 below the value for the non-suspended 2DEG. It is
likely that this strong reduction of electron mobility results from the
combined
effects of damage imparted by ion bombardment during the dry etch step, and
from inhomogeneous surface tension in the suspended structure (see Fig~2(a)).
This is currently under investigation.  For future samples the etching process
will be optimized in order to minimize damage to the device~\cite{tang}.
At low magnetic fields negative magnetoresistance is manifested, 
indicating the prevalence of weak localization or interaction phenomena.

In Fig.~3(a,b) are shown the magnetoconductance traces of two different
cavities
with open leads as depicted in the SEM micrographs in Fig.~2(b).
The devices constitute either
quantum cavities or quantum dots, depending on the coupling to the 
leads, i.e. number of transmitting channels $N$ in the leads with
a final conductance of $G = N 2 e^2/h$. We denote the two different structures
measured as cavity~A and~B, since the effective total resistance is far below
the resistance quantum $h/e^2$.
The measurements on the coupled dot device are discussed below (see Fig.~5).
The density obtained from the SdH-resonances at larger magnetic fields is 
$n_s \cong 5.5 \times 10^{15}$ m$^{-2}$ (cavity~A) and $n_s \cong 6.8
\times 10^{15}$
m$^{-2}$ (cavity~B). Here we do not take into
account the spin degeneracy factor, since the spin splitting only emerges
at $B \cong 4$~T (cf. arrows in Fig.~3(a,b)). The resulting
mobilities are $\mu_{cav A, cav B} =$ 0.4~m$^2$/Vs and 0.3~m$^2$/Vs,
strongly reduced by the barriers introduced (compared to the 'free' 2DEG).
As seen, the 
data at larger fields exhibit the common signature of a 2DEG, while 
at lower fields an intricate resonance structure is found. 

In focusing on this fine structure around $B = 0$~T, shown in Fig.~4, we 
are able to extract some of the properties of the electron gas in the
suspended membrane: The low temperature measurement at 4.2~K yields 
two distinct features, namely the coarse modulation of the resistance and a 
fine structure with a modulation period of 
$\delta B_{cav A} = (26.1 \pm 1.0)$~mT. This fine structure vanishes 
rapidly upon increasing the temperature (see traces for 10~K in Fig.~4(a,b)). 
Under illumination the carrier density varies
and the $B$-periodicity disappears, even at 4.2~K (Fig.~4(a)). 

Evaluating the periodicity of the fine structure by treating it as
Aharonov-Bohm (AB)
oscillations, we obtain a radius of the first cavity of 
$r_{AB}^{cav A} = \sqrt{\phi_0 /  (\pi \times \delta B_{cav A}) }  = $ 
$\sqrt{4.14 \times 10^{-15} \mbox{Tm}^2 / (\pi \times 20 \mbox{mT} )}
\approx 225$~nm and
with $\delta B_{cav B} = (26.3 \pm 0.5)$~mT for the second one $r_{AB}^{cav
A} = 220$~nm. 
With this value we obtain an 'electronic' diameter of the cavity of 450~nm
for the first ($A$) and 440~nm for the second cavity ($B$).
Compared to the lithographically defined size of 800~nm, we thus find 
a depletion depth of $\sim $~175~nm from the edges. 
The fact that AB-like oscillations are found without an opening in the center
could possibly be explained by some local inhomogeneity within the dot, e.g. 
local strain in the 2DEG, which might result in an effective depletion of the 
electrons in the center of the cavity. 
This also agrees with the disappearance of AB oscillations after
illumination (higher density). Apart from the pecularity of the existence
of AB-oscillations in such a sample it has to be noted that dissipation 
phenomena in free-standing 2DEGs seem to be less pronounced than in comparable
2DEGs connected to the bulk GaAs-crystal. 

The electronic radius found from the AB like oscillations is
confirmed by the coarse modulation, which can be explained by geometric
resonances
found earlier in strongly modulated one-dimensional electron 
systems~\cite{thornton89:2128,muller94:8938}. From this model we obtain 
the classical cyclotron radius $r_c = \hbar k_F /eB$ with $k_F = \sqrt{2 \pi
n_s}$ being the Fermi wave vector. For the two quantum cavities measured
(Fig.~3) we find a variety of different radii: For the first cavity~A
(Fig.~4(a)) 
we obtained
$r_c = 652$~nm, $r_c^d = 680$~nm, $r_c^c = 484$~nm, $r_c^b = 357$~nm,
$r_c^a = 265$~nm after illumination. Prior to illumination the only
cyclotron orbit
found is $r_c = r_c^c$. 
Similar commensurability oscillations and an almost identical fine structure
are found for cavity~B (Fig.~4(b)): $r_c^c = 874$~nm, $r_c^b = 539$~nm,
$r_c^a = 210$~nm.
The insets in Fig.~4(a,b) show some of the possible orbits: The upper 
sketch $\alpha$ shows the ideal
case, while in the lower one ($\beta$)
we allowed a finite contact width, demonstrating
the occurence of the different cyclotron radii. In the third
case $\gamma$ the electron follows a trajectory with maximum cyclotron
radius (with non-ideal contacts), elucidating the large values of $600 -
800$~nm measured. Considering the different radii and the corresponding
orbits we obtain for the smallest radii A:~$r_c^a = 265$~nm and B:~$r_c^a =
210$~nm
a reasonable agreement with the values obtained before under the assumption
of AB oscillations. 
Since we find similar radii for the commensurability oscillations, being
caused by the
coarse modulation, and under the assumption of AB-oscillations, caused by the
fine structure, we are confident that the depletion depth is correctly
determined.

Finally, we want to present some of the first transport data on {\it
suspended} quantum dots as shown in Fig.~2(b), i.e. for the case when
$G < e^2/h$. Already strong indications were found
that the relaxation mechanisms of single electrons tunneling through coupled
dots is influenced by 
discrete phonon modes~\cite{fujisawa,brandes}. 
However, these quantum dots were still embedded in the bulk GaAs crystal and
hence the phonon coupling is not well controlled. 
The conductance measurements on a quantum dot under bias voltage
strongly decoupled from the leads are shown in Fig.~5. 
For these measurements we employed the coupled dot system shown in
Fig.~2(b), since the smaller total
size and the narrower contacts result in an increased resistance.
As seen we find a transition into the tunneling regime for $B = 0$~T 
in the low drain/source bias ($V_{ds} < \pm 75$~mV) regime where the 
conductance $G = dI /dV$ is found to be below $e^2/h$. The conductance drops 
below $e^2/h$ for bias voltages around +20~mV to +50~mV. This drop is found
to occur stepwise. Including the magnetic field dependence we are able to
identify three different step widths, which are denoted $E_C^0$, $E_C^1$,
and $E_C^2$.
Interpreted as charging states of the quantum dots
these steps correspond to energies approximately
40~meV, 10~meV, and 30~meV, respectively. A crude estimation of the
charging energy
expected
from the electronic dot radius of roughly $r_e \sim 200$~nm (using the
values found above
for the depletion depth) we obtain $E_C = e^2/2C_{\Sigma} \sim 1$~meV. We
assumed a dielectric disk with 
$C_{\Sigma} = 8 \epsilon_0 \epsilon_r r_e$, where $C_{\Sigma}$ is the total
capacitance
of the dot, $\epsilon_r$ the dielectric constant of AlGaAs
and $\epsilon_0 = 8.85 \times 10^{-12}$~F/m.
Although this value is far below the stepsize found in Fig.~5,
it is still sufficient to allow Coulomb blockade (CB) at 4.2~K. Moreover, it 
is possible that the electronic diameter 
of the coupled dots is smaller than we estimated here. 

Unfortunately it was not possible to verify charging of the quantum dot
device by 
an external gate directly. Thus no clear CB signature 
can be identified for this device. Instead we applied a magnetic field 
perpendicular to the plane of the dot. Increasing the 
field, we observe a suppression and an additional modulation of the
conductance
around $V_{ds} = 0$, 
as seen for $B = 4$~T and 8~T. Discrete states in the quantum dot must
have formed, indicated by the structure in the $IV$-characteristic 
of Fig.~5. This is
evidenced by the modulation at finite magnetic field. 
Additionally, we find an enhanced background noise when keeping the
dots under forward bias, which can be attributed to an effective mechanical
vibration of the whole sample, due the resulting Lorentz force in the magnetic
field. 

In summary, we have demonstrated a technique of how to fabricate freely 
suspended two-dimensional electron gases. 
In transport measurements we observe the characteristic 
features of 2DEGs. Moreover, we find at low temperatures a 
strong negative magnetoresistance effect, indicating the lower mobilites of
the suspended electron gas compared to the bulk values.  
In further measurements on suspended quantum dots and quantum cavities we
found geometric resonances and a fine structure from which we estimate the 
lateral depletion length to be 175~nm.
Finally, we presented first measurements of transport spectroscopy on
a suspended quantum dot structure revealing the existence of discrete states. 
Although the properties of the suspended electron gas are not yet optimized
in comparison to conventional AlGaAs/GaAs-heterostructures, we clearly 
demonstrated functioning of these devices. Moreover, this technique offers a 
broad variety of possible applications, including high-frequency strain 
sensing in nano-mechanical devices, studies on the single electron/phonon
interaction, measurements of electronic specific heat 
and ultra-sensitive bolometry. 

% %%%%%%%%%%%%%%%%%%  
We gratefully acknowledge funding for this work from DARPA MTO/MEMS under
grant DABT-63-98-1-0012.  We thank A.N. Cleland for expert technical help
and S. Ulloa for critically reading the manuscript. RHB
thanks the Max-Planck-Gesellschaft (Otto-Hahn stipend), the
Max-Planck-Institute for
Solid State Physics, Stuttgart, and the Alexander-von-Humboldt Stiftung
(Feodor-Lynen stipend) for support. 

\newpage

$^{\dagger}$~current address: Center for NanoScience and
Sektion Physik, Ludwig-Maximilians-Universit\"at M\"unchen,
Geschwister-Scholl-Platz 1, 
80539 Munich, Germany. \\
{\it e-mail: robert.blick@physik.uni-muenchen.de} 

$^{\ddagger}$~new address: Institut f\"ur Angewandte und Experimentelle
Physik, Universit\"at Regensburg, 93040 Regensburg, Germany. 

$^{\amalg}$ IBM Research Staff Member Emeritius.

\newpage

Fig.~1:

(a) Calculation of the band-structure of the freely suspended two-dimensional
electron gas. As indicated the filled squares mark the non-illuminated trace
and the triangles the illuminated one. The Fermi level is pinned at mid-gap
for top and bottom surfaces. The temperature assumed in the calculations
is $T=$~10~K. Top and bottom of the heterostructure are formed by GaAs cap
layers, two AlGaAs:Si donor layers and two AlGaAs spacer layers around the
two-dimensional electron gas. \\

(b) Scanning electron beam micrograph of the suspended mini-Hall-bar used
in the 
measurements. The dimensions of this structure are length $\times$ width
$ = 2 \times 1~\mu$m$^2$, while the total thickness of the whole structure is
100~nm. \\

Fig.~2:

(a) Side-view of the device in Fig.~1(b), demonstrating the clear undercut;
for all
samples the same wafer material was used.\\

(b) Aerial view of the suspended single and double quantum dots. The 
lithographical diameter of the single cavity is 800~nm, while the double dots 
are on the order of 400~nm. We fabricated several of these samples; data
shown in Fig.~3,4 is obtained from two different cavities~A and~B, while
the data from
the coupled dot is presented in Fig.~5.\\

(c) Shown is the longitudinal magnetoresistance $R(B)$ of one of the
suspended 
mini-Hall-bars at $T=4.2$~K prior and after illumination as indicated (second
sample's
data not shown). The inset gives a $(1/B)$-plot demonstrating the periodicity 
of the oscillations after illumination (electron density for samples used is 
obtained as: $n_s \cong 5.5 \times 10^{15}$~m$^{-2}$). \\

Fig.~3:

(a) Magnetoresistance trace of the suspended 'open' quantum
dot (cavity~A), prior to illumination (note the large resistance). 
The leads contain several transport channels, resulting in 
a strong coupling, i.e. $G > 2e^2/h$. In the high field region the
classical Shubnikov-de~Haas
oscillations appear, while at low fields we find commensurability oscillations
(circle). The two arrows mark the onset of the spin splitting. Inset gives a 
clear $1/B$-periodicity of the oscillations, we obtain a density
of $n_s^{cavA} = 5.65 \times 10^{15}$~m$^{-2}$.\\

(b) Corresponding Shubnikov-de~Haas data of cavity~B
(evaluation of the data in the inset 
gives $n_s^{cavB} = 6.38 \times 10^{15}$~m$^{-2}$). For this cavity we find 
even better pronounced commensurability oscillations (circle). Inset
depicts the experimental setup for the transport measurements. Also marked
by arrows is the occurence of the first spin splitting. \\

Fig.~4:

(a) At the lowest temperatures (4.2~K-plot) we find commensurability
resonances and superimposed
$B$-periodic Aharonov-Bohm type oscillations from which we 
derive an 'electronic' diameter of the cavity of 
$\sim$~450~nm for cavity~A. The oscillations clearly vanish at 10~K and
under illumination.
The geometric resonances are also modified under illumination.  
The commensurability resonances equally allow the determination of the 
cavity diameter: Calculating the cyclotron radii we find 
$r_c = 652$~nm, $r_c^d = 680$~nm, $r_c^c = 484$~nm, $r_c^b = 357$~nm,
$r_c^a = 265$~nm (see text for details). 
Insets give some of the possible orbits: The upper 
sketch $\alpha$ shows the ideal
case, while in the lower one $\beta$ 
we allowed a finite contact diameter, demonstrating
the occurence of the different cyclotron radii. \\

(b) Similar commensurability oscillations and almost identical fine structure
for cavity~B.
Calculating the radii of the cyclotron orbits yields:
$r_c^c = 874$~nm, $r_c^b = 539$~nm, $r_c^a = 210$~nm.
Inset ($\gamma$) sketches a possible electron trajectory with 
maximum cyclotron radius. \\

Fig.~5: 

Drain/source conductance of the coupled dot in the tunneling regime for
different magnetic fields at 4.2~K. Marked with $E_C^0$, $E_C^1$, 
and $E_C^2$ are the different possible charging states, changing
upon increasing the field (see text for details).
Inset indicates sequential electron tunneling through the coupled dot. \\

\end{document}